# New Security Challenges Towards In-Sensor Computing Systems

Mashrafi Kajol and Qiaoyan Yu
*Department of Electrical and Computer Engineering, University of New Hampshire*
Durham, NH 03824, USA

*Abstract*—Data collection and processing in advanced health monitoring systems are experiencing revolutionary change. In-Sensor Computing (ISC) systems emerge as a promising alternative to save energy on massive data transmission, analog-to-digital conversion, and ineffective processing. While the new paradigm shift of ISC systems gains increasing attention, the highly compacted systems could incur new challenges from a hardware security perspective. This work first conducts a literature review to highlight the research trend of this topic and then performs comprehensive analyses on the root of security challenges. This is the first work that compares the security challenges of traditional sensor-involved computing systems and emerging ISC systems. Furthermore, new attack scenarios are predicted for board-, chip-, and device-level ISC systems. Two proof-of-concept demos are provided to inspire new countermeasure designs against unique hardware security threats in ISC systems.

## I. INTRODUCTION

Advanced health monitoring systems demand fast data collection and processing. Traditionally, massive sensor data are continuously collected to support external processing like Artificial Intelligence (AI)-based analysis and prediction [1], [2], which consumes significant power and introduces delays. To address this limitation, wearable health-monitoring devices need to be power-efficient, compact, and capable of processing large amounts of data in real time without sacrificing accuracy [3]. One promising direction is to enable sensing units [4]–[6] to offer additional computing capability [7], [8] than a simple sensor.

In-Sensor Computing (ISC) emerges as a new computation paradigm [9] to address the increasing concern on latency and energy consumption in sensory data transmission, analog-to-digital conversion (ADC), and data pre-processing. Sensing units in an ISC do not only sense the target surroundings but also process data at the point of collection, rather than requiring extensive data transfer [9]. A conceptual diagram of in-sensor computing systems is shown in Fig. 1(a). Sensors have some computation power, the essence of which is a desired processing function implemented by leveraging specific material/device mechanisms. Traditional ADC and logic functions are omitted from the data path before the main digital computation unit. The local processing capability ensures real-time monitoring, which is especially critical for health applications such as glucose monitoring, cardiac health tracking, and emergency response systems [10], [11].

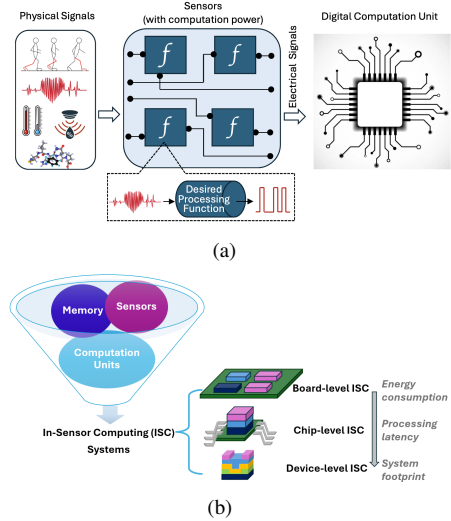

Figure 1. A conceptual diagram for in-sensor computing. (a) Overview and (b) technology integration.

Essentially, an in-sensor computing system integrates sensors, memory elements, and computation units into one system, as shown in Fig. 1(b). Depending on the integration technology, ISC can be implemented at board, chip, and device levels. As the system footprint decreases from board level to device level, the corresponding energy consumption and analog signal processing latency also decrease because of less ADC conversion and sensory data transmission/receiving.

Existing literature for ISC mainly investigates the materials and device structure to implement desired functions, pursue better performance [12], [13], and examine the feasible integration technologies [7], [14]. While offering significant benefits in performance improvement and power saving [15]–[17], in-sensor computing could also bring new security challenges. Unfortunately, limited work is available to study new threats and unique attack surfaces for ISC systems. This work fills the gap by making the following contributions:

- Comprehensive analyses of the attack scenarios in ISC systems are performed to characterize the new and unique attacks. To the best of our knowledge, this is the first work that analyzes security challenges in ISC systems.
- Proof-of-concept demos are presented to guide the development of potential countermeasures against hardware security threats in ISC systems.

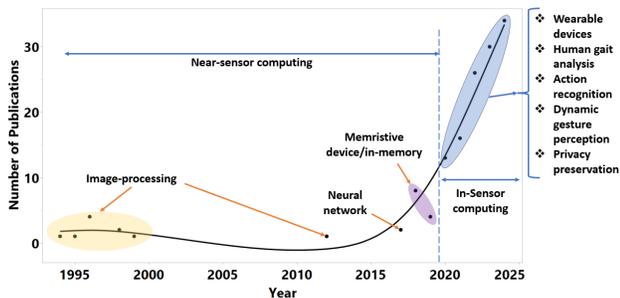

Figure 2. A historical progress of in-sensor computing.

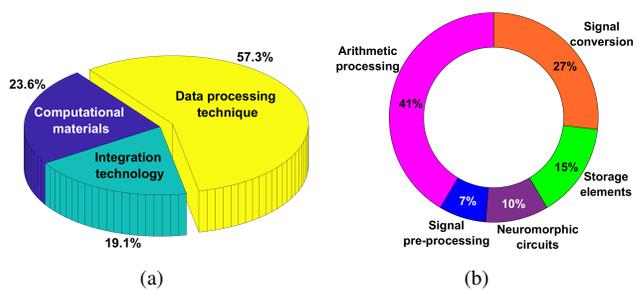

Figure 3. In-sensor computing publications from 2020 to 2024 (Sept.) (a) a pie chart for the percentage of publications with three distinct categories (b) a donut chart for specific feature block-related publications in the data processing technique area.

## II. SURVEY OF IN-SENSOR COMPUTING

### A. Prior to In-Sensor Computing

Over the past few years, the field of sensor computing has experienced a noticeable shift. The overall trend from 1994 to 2024 (Sept.) is shown in Fig. 2. The majority of research before 2020 was conducted on near-sensor computing, where computation was performed in close proximity to sensors. As a result, there has been an increase in data processing efficiency and a decrease in communication latency between sensor and computing units. As data-intensive applications demand more and more effective and real-time data processing, there is a significant rise in publications about in-sensor computing after 2020. Ever since then, the popularity of ISC systems has steadily grown because ISC's advantages in speed, functionality, and energy efficiency are attractive for applications such as wearable devices, autonomous systems, health monitoring, and AI-driven image recognition.

### B. In-Sensor Computing (ISC) Systems

Based on the underlying focus, we categorize the existing in-sensor computing literature into three branches: computational materials, integration technology, and data processing techniques, as shown in Fig. 3(a). The distribution of research articles on in-sensor computing indicates that data processing techniques take 57.3% share of the total publications.

**Computational materials-based ISC** leverages novel materials with intrinsic computational capabilities, such as phase-change materials and 2D materials like transition metal dichalcogenides. These materials are designed to sense and process data simultaneously. For example, perovskites-based photoelectric materials [12] can process visual information directly in hardware, eliminating the need for separate computational units.

**Integration (fabrication) technology-based ISC** focuses on how different components (sensors, processors, and memory units) are fabricated and integrated to perform computation at a sensing node. Technologies such as 3D integration of CMOS sensors and memory arrays enable efficient data flow and real-time processing without needing an external processing unit. The 3D stacking of dynamic vision sensors with an in-sensor processing layer [14] helps to compress data and reduces latency in machine vision applications.

**Data processing technique-based ISC** exploits special design and optimization of algorithms and architectures to enable efficient processing of sensor data. For applications such as pattern or image recognition, techniques like neuromorphic computing, reservoir computing, and convolutional neural networks (CNNs) have been applied to expand the functionality of sensors. For instance, the work [18] implements reservoir computing in optical sensors to process data for gesture or object recognition with high accuracy and reduced computational complexity. In Fig. 3(b), we zoom in on the research progress made in data processing techniques for ISC systems. The dominant focus areas are arithmetic processing (41%) and signal conversion (27%). Mathematical operations (e.g., multiplication and accumulation) are redesigned for sensors [8]. Signal conversion techniques use new quantization or encoding to transform sensory data into a form easier to process or transmit [19]. Other data pre-processing [20] and in-memory computing [21], [22] methods are also interested in the ISC community.

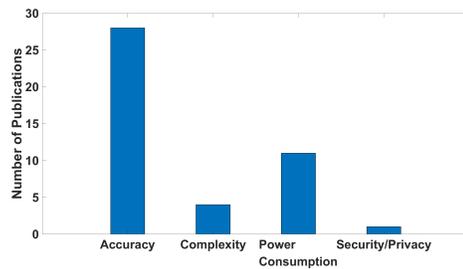

Figure 4. A bar chart for primary focuses of the existing publications.

### C. Unexplored Aspect of ISC Systems

The design metrics discussed in the existing publications for ISC systems include accuracy, complexity, power consumption, and security/privacy. As shown in Fig. 4, most of the existing works [16], [23], [24] pursue high accuracy in data processing. A good amount of work [15]–[17] explores power-saving techniques while maintaining accuracy. A few works [25] aim to reduce the complexity of processing units. As the security/privacy aspect is unexplored [26], this work fills this gap by analyzing new attack threats and suggesting potential countermeasure designs.

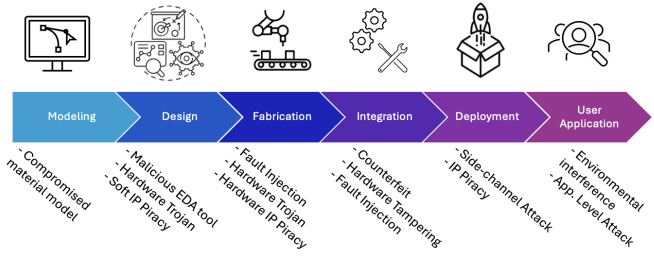

Figure 5. Attacks in different phases of ISC system's design and deployment.

Table I
COMPARISON OF ATTACKS IN TSC AND ISC SYSTEMS

| Systems | TSC | ISC |
|---|---|---|
| Attack Accessibility | **High:** through interfaces between sensors and computation units or remote access | **Limited:** physical access to the entire system, not directly connection to individual components |
| Prior Knowledge | Require **partial** knowledge of target components | Require **full** knowledge of entire system |
| Attack Tools | **Low:** many existing tools available to automate attacks | **High:** customized tools are needed |
| Attack Detection | **Easy:** probe via multiple wire/wireless interfaces and use existing validation tool | **Complicate:** lack generic validation tools and need to localize/differentiate attack sources |

## III. PROPOSED ATTACK SURFACE ANALYSES

### A. New Features of Attacks in ISC Design Flow

Attacks in ISC systems can be performed at every stage of the design flow, as shown in Fig. 5. Since in-sensor computing heavily relies on the special mechanism offered by materials to perform sensing and computation, material modeling plays a vital role in the design flow of ISC systems. As a result, the material and device modeling phase is the first vulnerable stage. Compared to Traditional Sensor-involved Computing (TSC) systems, it is more critical for ISC to ensure the trustworthiness of sensor models and design verification tools. This is because all functionalities, including sensing and computation, of the system are determined and tested at the design phase. The confidentiality of the system design is more centralized in ISC than TSC. In the fabrication stage, a malicious foundry can pose more security threats to ISC systems than TSC systems because the foundry can tamper with both analog and digital components of the system. Typical attacks such as fault attacks, side-channel attacks, and hardware Trojan insertion observed in TSC systems could also challenge the integrity of ISC systems. Table I provides more comparisons of different aspects of attacks in TSC and ISC.

### B. Attack Surfaces at Different ISC Levels

*1) Attacks at Board-Level ISC:* From the board level, the digital memory elements and computation units will suffer from the same traditional attacks on hardware. New attacks at this level are highlighted in Fig. 6(a). Because sensors are the core units of ISC, changing the sensor's surrounding environment (humidity, temperature, pressure, etc) in a confined local region will directly interfere with the sensory data before computation. This type of attack only needs preliminary knowledge of the ISC functionality. In new side-channel attacks, the data transmission between sensors (S) and memory elements (M) may not be in a digitized format.

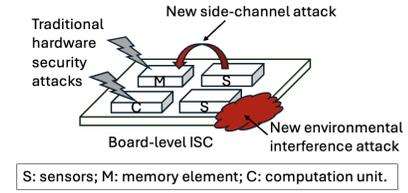

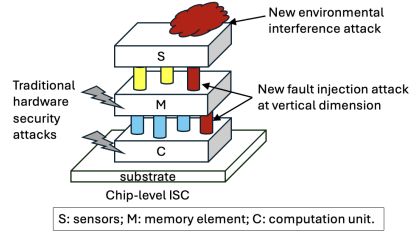

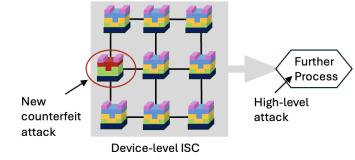

Figure 6. Attack surfaces at (a) board-level (b) chip-level (c) the device-level ISC systems.

The analog or semi-analog signal from sensors, especially bio-sensors, may incur new side-channel attacks, for instance, the attack intending to leak privacy. The analog format may reveal more details than the digital format since analog signals have frequency, magnitude, and phase.

*2) Attacks at Chip-Level ISC:* Sensor nodes will be deployed to public hostile locations, where sensor devices are vulnerable to physical attacks. We could not completely isolate or shield the systems from the sensing environment (otherwise the system could not function as it is supposed to do). Figure 6(b) shows a stack-based three-dimensional (3D) ISC system and highlights the potential attacks that could occur in chip-level ISC systems. As the system is expanded from the vertical dimension, malicious integration entities could introduce fault attacks to the vertical chip-to-chip interconnect. Depending on the specific materials used in the 3D integration, fault characteristics for those attack scenarios should be studied case-by-case. The non-generalized analog-fault attack model leads to new challenges for ISC systems.

*3) Attacks at Device-Level ISC:* If all sensing and computation are performed at the device level as shown in Fig. 6(c), the implementation of attacks needs more knowledge, advanced tools, and fine-tuning than the attacks at the board- and chip-level ISC systems. At device-level ISC systems, customized counterfeit attacks could be a new threat, which needs the full understanding of sensors. It would be a white- or grey-box attack. Another attack surface would be the later stage of the ISC system, which continues the computation after the pre-processing at the device level.

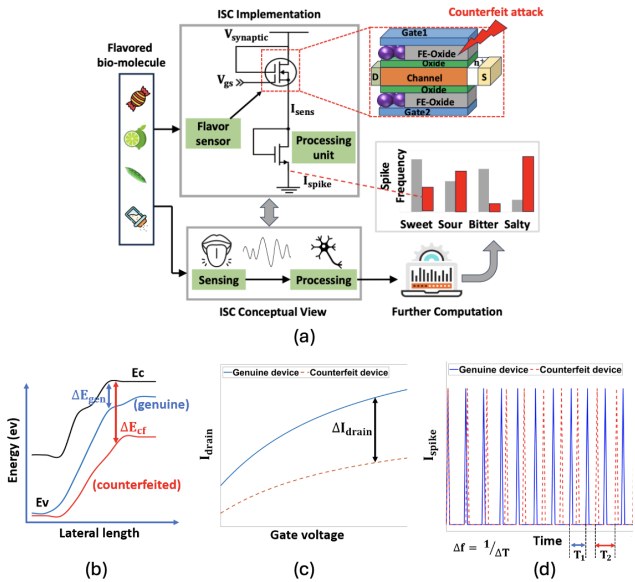

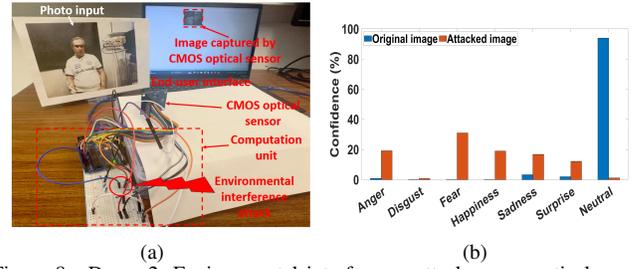

(a)                  (b)

Figure 8. Demo 2: Environmental interference attack on an optical-sensor-based ISC for facial-expression-based emotion recognition. (a) Experimental setup for an ISC system for emotion detection and (b) impact of temperature interference on detection accuracy.

Figure 7. Demo 1: counterfeit attack on an ISC for biomolecule detection. (a) Experimental setup, (b) changes in energy band (conceptual), (c) changes in I-V characteristics, and (d) changes in detection output (frequency).

## IV. DEMO EXAMPLES OF ATTACKS IN ISC SYSTEMS

### A. Counterfeit Attack at Device-Level ISC

This section demonstrates the means and consequences of a counterfeit attack in an ISC-based gustatory system [27], which is utilized in a wide range of applications including pharmaceutical, medical, and food/beverage industries. The experimental setup is shown in Fig. 7(a). The double gate ferroelectric tunnel FET (DG-FE-TFET) is used as a sensor device and a single transistor-based computing unit is utilized to mimic spiking behavior. We assume that a counterfeit material was used in the ISC fabrication stage to alter the energy band and thus the I-V characteristic of the sensing device. Due to the high integration density and the lack of probing locations, a small change in the FE-Oxide thickness of DG-FE-TFET was not detected until the application completed the massive data processing. As shown in Fig. 7(d), the offset frequency induced by the wrong $I_{spike}$ leads to a wrong biomolecule recognition. This demo indicates that the detection of counterfeit attacks at the device-level ISC will be delayed due to the lack of probing opportunities for early diagnosis.

### B. Environmental Interference Attack at Board-Level ISC

The proof-the-concept example in this section is used to demonstrate a board-level ISC system is vulnerable to a low-cost environmental interference attack. This ISC system captures a facial expression image via a CMOS optical sensor [28] and then utilizes the Py-Feat machine learning model [29] to classify emotions. The experimental setup is shown in Fig. 8(a). We assume that an attacker only knows the overall function of the ISC system but lacks knowledge of the ISC implementation details. A low-cost malicious circuit composed of a photo-resistor, a transistor, and a regular resistor was mounted on the serial data bus between the sensor and the computation unit. An illumination-adjustable light was used to influence the resistance of the photo-resistor and thus sabotage the integrity of the captured images. Figure 8(b) shows that the environmental interference attack altered the classification conclusion (from neutral to five other emotions). This demo indicates that the resilience of ISC systems against harsh environments will be increasingly important than TSC systems. Due to a lack of authentication between sensors and computation units in ISC, the verification of data integrity in ISC is more challenging than in TSC.

## V. CHALLENGES AND FUTURE DIRECTIONS

Although ISC systems address the bottleneck in the communication from sensors to computation units, more analog signals are involved in ISC systems than traditional sensor systems. Consequently, existing countermeasures designed for digital systems may not be effective in thwarting new fault attacks, which could be disguised as analog noise. The high integration density of ISC systems makes it more difficult to probe intermediate checking points for early attack diagnosis. Moreover, it is not feasible to authenticate individual sensing components and computation units. As ISC systems exploit various materials for different sensing purposes, it is challenging to develop generic design validation tools to build systems' equivalent models and examine the integrity of system design and implementation. The inherent nature of parallelism in some ISCs (i.e., image processing, memristive array) demands a new screening process to check all ISC elements. To address those new challenges, research efforts are required in various sectors to maintain data integrity and develop lightweight encryption algorithms tailored for analog processing. Besides, a hybrid countermeasure may be useful to combine small external validation and robust internal defense.

## VI. CONCLUSION

In-sensor computing systems present a promising solution for various applications in the healthcare sector to achieve high processing speed and low energy cost. However, hardware security challenges of in-sensor computing systems have not been widely investigated. This work fills this gap by analyzing security threats from board, chip, and device levels. We hope that our proof-the-concept demos will inspire researchers to make efforts to strengthen the security of ISC systems.